# Interlayer Thermal Conductivity of Rubrene Measured by ac-Calorimetry

H. Zhang and J.W. Brill

*Department of Physics and Astronomy, University of Kentucky, Lexington, KY 40506-0055, USA*

We have measured the interlayer thermal conductivity of crystals of the organic semiconductor rubrene, using ac-calorimetry. Since ac-calorimetry is most commonly used for measurements of the heat capacity, we include a discussion of its extension for measurements of the transverse thermal conductivity of thin crystals of poor thermal conductors, including the limitations of the technique. For rubrene, we find that the interlayer thermal conductivity, $\approx 0.7$ mW/cm·K, is several times smaller than the (previously measured) in-layer value, but its temperature dependence indicates that the interlayer mean free path is at least a few layers.

## I. INTRODUCTION

Rubrene, $C_{42}H_{28}$, is a layered crystalline organic semiconductor which is of considerable current interest because of its relatively high electronic mobility, e.g. $\mu \sim 8$ cm$^2$/V·s.[1] The rubrene molecule consists of a tetracene backbone with four phenyl side groups. In the crystal, the tetracene groups stack in the high conductivity *ab* plane with the phenyl groups sticking out of the plane along **c**.[2,3] The in-plane thermal conductivity near room temperature has been measured to be $\kappa \sim 4$ mW/cm·K,[4] a value similar to many other van der Waals bonded organic crystals[5] as well as polymers.[6,7] While such low values of $\kappa$ may limit the use of these organic semiconductors in microelectronics, they are favorable for possible thermoelectric applications.[7]

Because rubrene has a layered structure, it is also important to know its interplane thermal conductivity. Since crystals are typically < 0.2 mm thick and the thermal conductivity is low, it is difficult to use conventional techniques for such measurements, as there can be comparatively large corrections for both thermal radiation and heat transported by leads. A number of frequency dependent techniques, in which oscillating power is applied to the sample and resulting temperature oscillations measured, have been developed to measure thermal conductivity of small samples.[7-15] For these, the power transported out of the sample by leads and radiation determines the "external" thermal time constant, $\tau_1$, but measurements are typically made at frequencies $\omega \gg 1/\tau_1$,[8] so that heat loss and radiation corrections are negligible.[11]

Perhaps the most widely used of ac-methods is the "3$\omega$" technique,[11] in which oscillating power is applied to a Joule heater deposited on the sample and the temperature oscillations of the heater measured. One disadvantage of the 3$\omega$ technique is that for a layered, anisotropic crystal the heater's oscillating temperature will be a function of the geometric mean of the "in-plane" and transverse thermal conductivities,[15] so the two components are not separately determined; in addition, it may be difficult to apply a low-noise Joule heater to some organic materials without damaging the material. Alternatively, other investigators have used chopped light to apply oscillating power to one surface of the sample and measured the oscillating temperature at different positions and/or frequencies;[6-10, 12-14] the frequency dependent response is a function of the thermal diffusivity, $D \equiv \kappa/c\rho$, where c is the specific heat, $\rho$ the mass density, and $\kappa$ the thermal conductivity. Most commonly, the oscillating temperature is measured on the opposite surface as a function of the lateral distance from the light, either by screening the light from part of the sample[8-10,14] or by using a laser to illuminate a small spot.[14] By working at frequencies $\omega \ll 1/\tau_2$, where

$$\tau_2 \equiv d^2/\sqrt{90}D_{tran} \qquad (1)$$

is the "internal" thermal time constant determining how quickly heat gets through the sample of thickness d and $D_{tran}$ is the transverse diffusivity (i.e. parallel to d),[16] these techniques determine the *in-plane* thermal diffusivity. However, Kato, *et al*, have also measured the transverse diffusivity by measuring the frequency dependence of the phase of the oscillating temperature at a point directly behind a small illuminated spot.[14]

To measure the temperature dependent transverse thermal conductivity, $\kappa_c$, of rubrene, we have used an alternate technique in which instead of illuminating a small spot, the entire face of the sample is uniformly illuminated, and the frequency dependence of the *magnitude* of the temperature oscillations on the opposite surface are measured. The technique is an extension of the well-known ac-calorimetry technique used to measure the heat capacity of small samples,[16-18] and has the advantage of simplicity of set-up and sample mounting, useful for small and fragile samples, but has some important limitations. We will therefore begin by describing our technique and its limitations in some detail.

## II. EXPERIMENTAL TECHNIQUE

We consider a sample with total heat capacity (including any addenda) C, area A($\approx$Lw) and thickness d illuminated on its top surface by light of intensity $P_0$ chopped at frequency $\omega$. If the sample is sufficiently thin[13,19] (e.g. d<<L,w) and the light is absorbed uniformly on the surface, the heat flow will be one-dimensional and, if $\omega >> 1/\tau_1$, the magnitude of the oscillating temperature on the back surface is given by[16]:

$$T_\omega = 2P_0 A \Phi(\omega)/(\pi\omega C) \quad (2)$$
$$\text{with } \Phi(\omega) = \sqrt{2}\chi/[\sinh^2\chi \cos^2\chi + \cosh^2\chi \sin^2\chi]^{1/2} \quad (3)$$
$$\text{and } \chi \equiv [\sqrt{90}\, \omega\tau_2/2]^{1/2} = [\omega d^2/2D_{tran}]^{1/2} = [c\rho d^2/2\kappa_{tran}]^{1/2} . \quad (4)$$

In the limit $\omega\tau_2 << 1$,
$$\Phi(\omega) \approx 1/[1+(\omega\tau_2)^2]^{1/2} \quad (5),$$

so that $T_\omega \alpha\ 1/C$, and this is the well-known ac-calorimetric technique for measuring the heat capacity.[16-18] At higher frequencies, the frequency dependence of $T(\omega)$ determines $\tau_2$ and hence the transverse diffusivity. The exact and approximate expressions for $\Phi(\omega)$ (Eqtns. (3) and (5)) are plotted in the inset to Figure 1; the approximate expression actually still holds well at $\omega\tau_2 \sim 1$ (e.g. $\approx$ 2% error). Because $\tau_2$ depends quadratically on d, it is necessary for the thickness to be well-known for a precise, absolute determination of $D_{tran}$ and $\kappa_{tran}$, but relative measurements (e.g. as a function of temperature) are possible even when it is difficult to determine the thickness, as shown below.

In our experiments,[18] we use light from a 100 W quartz-halogen lamp focused through a 1.6 mm diameter optical fiber onto the window of a cryogenic header. Within the header, the light travels through a silvered glass rod to provide a ~ 5 mm diameter light spot on the sample. A black, light absorbing PbS film (~ 30 nm thick) is deposited on the top face of the sample. A chromel-constantan thermocouple, prepared by spot-welding slightly flattened 25 μm diameter wires in a cross,[18] is glued to the center of the back of the sample with a thin layer (< 10 μm) of butyl acetate based silver paint.[20] In a typical vacuum of ~ 0.1 torr with the sample simply supported by the four thermocouple wires, the light typically increases the average (dc) temperature of the sample by < 10 K with $T(\omega)$ ~ 0.2 mK at 100 Hz. The oscillating thermocouple voltage, $V_\omega$, is measured with a two-phase lock-in amplifier with a transformer input and calibrated frequency dependence.[21] When the signal at high frequencies is very small (e.g. < 10 nV), the lock-in is operated in "XY" mode so that "no light" offset signals can be subtracted.

Because we are measuring the frequency dependence of the oscillating signal, it is important to consider the response time of the thermocouple and attaching glue. Figure 1 shows the frequency dependence of $fV_\omega$ ($f \equiv \omega/2\pi$) for a thermocouple soldered directly on to a d=130 μm copper foil with intrinsic $\tau_2 \approx$ 15 μs; it is seen that the response is independent of frequency to at least 200 Hz. However, the response time for samples with thermometers attached with the silver paint are significantly slower;[22] two examples are shown in Figure 1, with typical measured response times (i.e. from fits to Eqtn. (5)) of $\tau_{meas}$ ~ 1 ms. Because of the high silver content and resulting high thermal conductivity (~ 40 mW/cm·K)[20] of the paint, most of this delay is presumably due to the interface thermal resistance, $R_{ifc}$, between the paint and sample; i.e $\tau_{meas} \sim \tau_{ifc} = R_{ifc}C_{add}$, where $C_{add}$ is the heat capacity of the silver paint[23] and thermocouple[24] addenda. For $\tau_{ifc}$ ~ 1 ms and contact area S ~ 1 mm$^2$, the interface thermal resistivity $R_{ifc}S$ would need to be ~ 0.1 cm$^2$·K/W, similar to the value reported for metal-filled epoxies on aluminum,[25,26] and, in fact, we have not found a "faster" glue. Assuming that $\tau_{meas} = \tau_2 + \tau_{ifc}$, the sample must have $\tau_2$ > 3 ms, assuming a similar value of $R_{ifc}$. Of course, one cannot always assume similar interface resistivities for the silver paint on different materials; indeed, we have observed $R_{ifc}S$ > 1 cm$^2$·K/W for the silver paint on ~ 100 μm thick plexiglass and teflon sheets. One can avoid the problem of interface thermal resistance by using a thin film bolometer deposited directly on the sample or by measuring the oscillating thermal radiation from the back of the sample, but the thermocouple has the advantage of simplicity of mounting and high signal/noise. For our rubrene samples, we argue below that $R_{ifc}S$ is small based on the temperature dependence of $\tau_{meas}$.

The conditions for one-dimensional heat flow require L,w < the effective diameter of the light spot; in our case we require that L,w < ~ 2 mm. If the absorbed intensity varies by ΔP over a length $\Delta L$ on the surface, e.g. due to an inadequate PbS film, one may expect $\Phi(\omega)$ to increase by $\sim \Delta P/P_0$ at frequency $\omega_{long} \sim (D_{long}/D_{tran})(d/\Delta L)^2/\tau_2$, where $D_{long}$ is the longitudinal (i.e. in-plane) diffusivity. An example is shown in Figure 2 for a relatively thick (d=550 μm), "A-plane", i.e. d // (11$\underline{2}$0), sapphire plate, where a small step is observed near 10 Hz. Nevertheless, the frequency dependence of $V_\omega$ for 10 Hz < f < 50 Hz agrees very well with the expected value of the transverse (*ab*-plane) diffusivity,[27] $D_{ab} = 0.102$ cm$^2$/s (i.e. $\tau_2 = 3.1$ ms), indicating that the interface thermal resistivity is < 0.1cm$^2$·K/W and demonstrating the validity of the technique.

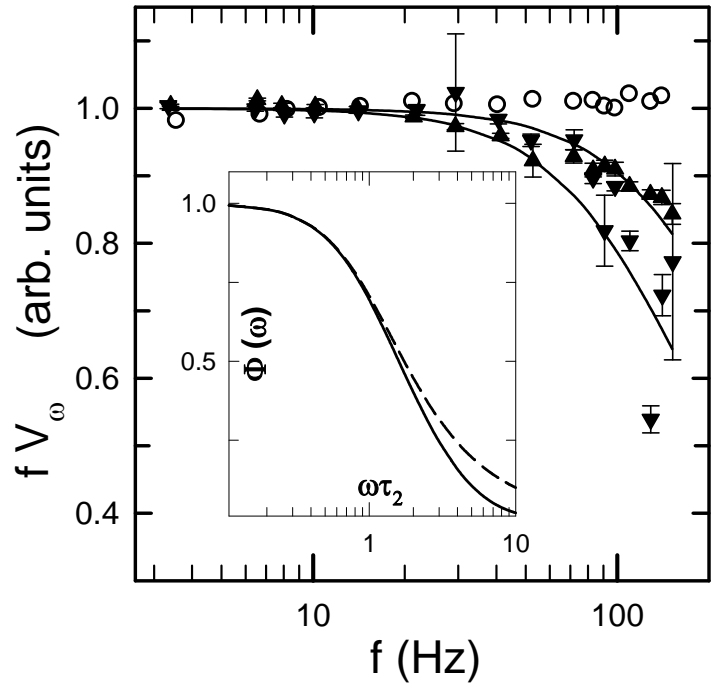

**Figure 1.** Frequency dependence of the room temperature thermocouple signals for 130 μm copper foils at room temperature. Open circles: thermocouple soldered directly to foil. Up and down solid triangles: thermocouples attached with silver paint. For reference, the curves show the responses corresponding to time constants of 0.75 ms and 1.25 ms. The inset shows $\Phi(\omega)$ calculated from Eqtn. (3) (solid curve) and Eqtn. (5) (dashed curve).

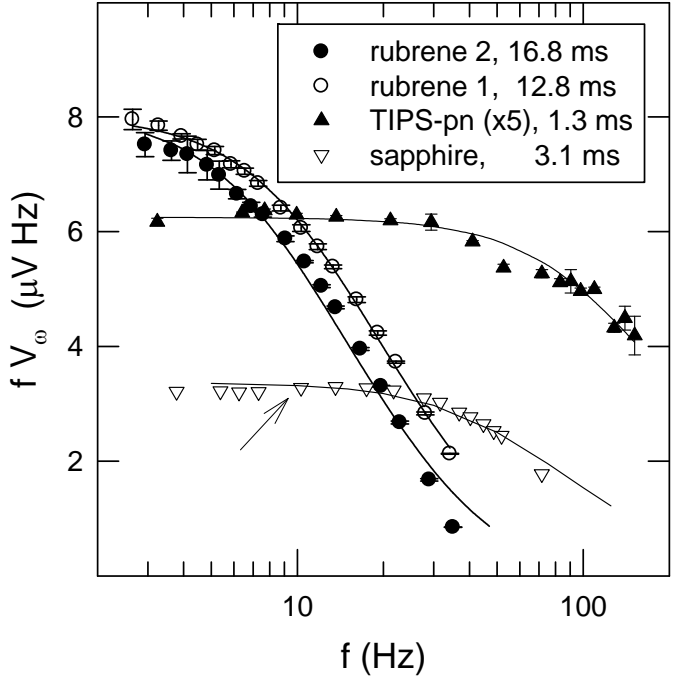

**Figure 2.** Frequency dependence of the room temperature thermocouple signals from two rubrene crystals, a TIPS-pn crystal, and a sapphire crystal. (For TIPS-pn, the data have been multiplied x 5). The curves show fits to Eqtn. (3); the time constants of the fits are shown in the legend. The arrow shows the small low frequency step associated with non-uniform heating for the sapphire sample.

### III. RUBRENE RESULTS AND DISCUSSION

Because the ac technique yields the transverse thermal diffusivity, it is necessary to know the specific heat to obtain the thermal conductivity. The inset to Figure 3 shows the temperature dependence of the specific heat of a rubrene pellet measured with differential scanning calorimetry (DSC)[28] with a precision of ~ 3%. Also shown on the scale on the right is the specific heat normalized to the gas constant, R, using the molecular mass M = 533 g/mole. Since there are 70 atoms/per molecule, the Dulong-Petit value of the specific heat is $c_P M = 210\,R$, so the room temperature specific heat is $\sim 1/3$ of its high temperature, saturated value. This presumably reflects the large number of non-propagating molecular phonons with energies greater than $k_B T$. Below, we assume that the room temperature specific heat associated with propagating acoustic phonons is saturated (or almost saturated) at its maximum value $c_P M \sim 3R$ ($c_P \sim 0.047$ J/g·K).

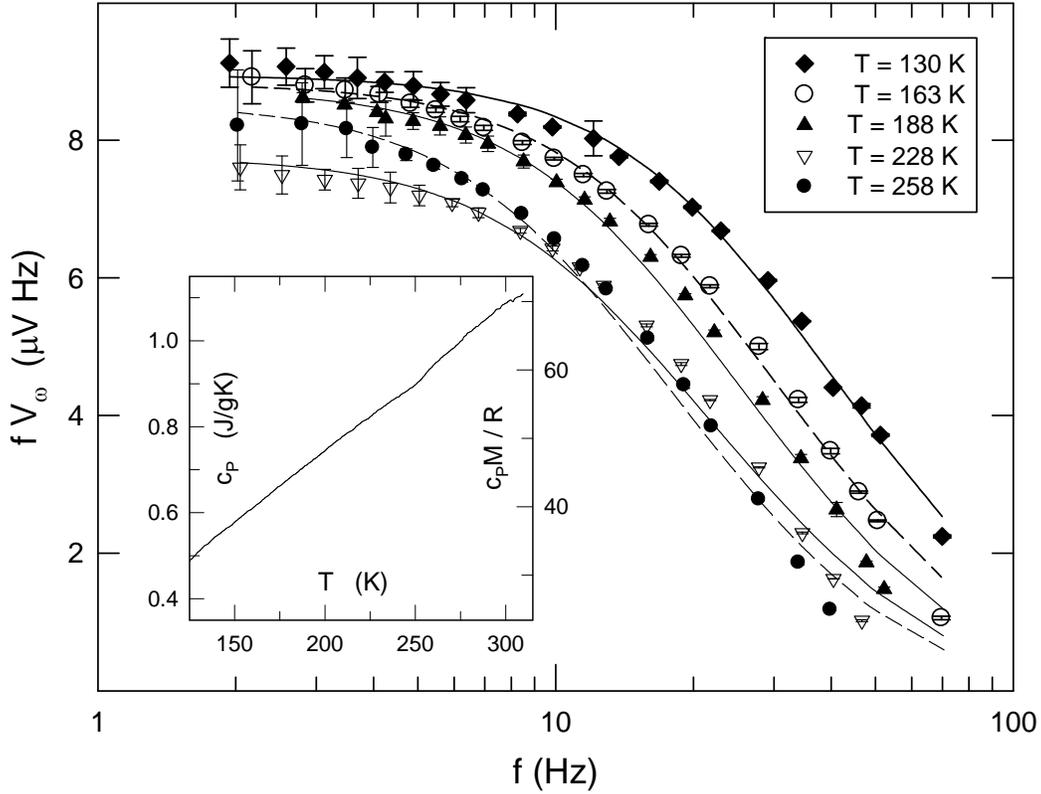

**Figure 3.** Frequency dependence of the thermocouple signals for rubrene crystal 2 at several temperatures. (Error bars at high frequencies are smaller than the points.) The curves show fits to Eqtn. (3); note the deviations at high frequencies at the higher temperatures presumably due to interface thermal resistance. The inset shows the temperature dependence of the specific heat measured by DSC for a rubrene pellet; the right scale shows the molar heat capacity normalized to the gas constant.

The frequency dependence of the thermocouple signal for two crystals at room temperature is shown in Figure 2. Although both crystals were very flat with smooth parallel faces, because these faces had irregular shapes and because the crystals are very soft and fragile, their thicknesses, 83 μm (crystal 1) and 90 μm (crystal 2), could only be estimated ± 10%. The figure shows the fits of $fV_\omega$ to Eqtn. 3, with time constants 12.8 ms and 16.8 ms (± 2%), so the calculated values of the transverse thermal conductivities for the two crystals, assuming $\tau_2 \approx \tau_{meas}$ and using density[3] $\rho = 1.26$ g/cm$^3$ and our measured specific heat, are $\kappa_c(1) = (0.78 \pm 0.20)$ mW/cm·K and $\kappa_c(2) = (0.70 \pm 0.18)$ mW/cm·K. The closeness of these results is one indication that the results are intrinsic (i.e. $\tau_2 \gg \tau_{ifc}$); a stronger indication is the temperature dependence of $\tau_2$, discussed below. (If instead we assume that $\tau_{ifc} \sim 1$ ms for both samples, the values of thermal conductivity would be $\kappa_c(1) = 0.85$ and $\kappa_c(2) = 0.74$ mW/cm·K.)

$\kappa_c$ is therefore almost an order of magnitude smaller than $\kappa_{ab}$.[4] The small values of both the in-plane and interlayer thermal conductivities make rubrene an attractive material for thermoelectric applications,[29] if doped samples can be prepared that keep a relatively high electronic mobility.

The fact that $\kappa_c << \kappa_{ab}$ suggests very poor energy transport via the phenyl side groups linking neighboring layers. If we assume that the acoustic phonon mean-free-path $\lambda \sim$ the interlayer spacing (1.4 nm) (Ref. 2) and that the acoustic phonon specific heat is close to its Dulong-Petit value, as discussed above, then this would give an interlayer acoustic phonon velocity

$$v_c \approx 3\kappa_c/(c_{acoustic}\rho\lambda) \approx 3 \times 10^5 \text{ cm/s}, \qquad (6)$$

which is actually a value fairly typical for organic molecular crystals.[30,31] Since $v_c$ is probably somewhat smaller than this, reflecting the weak interlayer coupling, the mean-free-path must instead be at least a few layers.

Our preliminary measurements on layered organic semiconductors with other side groups have indicated much larger values of interlayer thermal conductivity. Figure 2 also shows our results for a d $\sim$ 335 μm thick crystal of 6,13-bis(triisopropylsilylethynyl)-pentacene (TIPS-pn).[32] In the crystal, the pentacene backbones of the molecule form a brick-layer structure in the *ab* plane with the silyl side-groups sticking out of the plane.[32] The effective time constant for this sample is 1.3 ms and we cannot assume that this is intrinsic, i.e. not affected by interface thermal resistivity which would again only need to be $\sim 0.1$ cm$^2$·K/W to account for the thermal response. (Unfortunately, thicker crystals, needed for a slower response, have not been available.) Therefore, we can only estimate a lower limit for the transverse thermal conductivity; using density[32] $\rho = 1.1$ g/cm$^3$ and room temperature specific heat, measured by DSC (not shown), c = 1.48 J/g·K, we obtain $\kappa_c > 140$ mW/cm·K, indicating excellent thermal conduction between the silyl side groups. (Alternatively, the relatively thick TIPS-pn crystals we checked may have had stacking faults so that the interlayer thermal resistance was "short" by the in-plane conductance.)

Plots of the frequency dependence of $fV_\omega$ for rubrene crystal 2 at several lower temperature, with fits to Eqtn. (3) are shown in Figure 3. Figure 4 shows the temperature dependence of the measured time constant, which decreases roughly linearly with decreasing temperature, falling to $\sim 1/3$ of its room temperature value at T = 130 K. The specific heats of silver paint,[23] chromel,[24] and constantan[24] all decrease by only $\sim 30\%$ between room temperature and T = 130 K. Therefore, if the time constant is primarily due to the interface thermal resistance, its temperature dependence would imply that $R_{ifc}$ *decreases* with decreasing temperature, opposite to what is usually observed.[26] We therefore assume that the time constant is (at least) mostly intrinsic, i.e. $\tau_2 > \tau_{ifc}$.

Figure 4 also shows the temperature dependence of the interlayer thermal conductivity, normalized to its room temperature value, using the measured values of $\tau_{meas}$ and specific heat and assuming that corrections due to thermal expansion are negligible. Results are shown assuming that $\tau_2 = \tau_{meas}$ (solid triangles) and $\tau_2 = \tau_{meas} - 1$ ms (open triangles). (The interface thermal resistance typically increases by $\sim 50\%$ between room temperature at T = 130 K.[26] Since the addendum heat capacity falls by $\sim 30\%$, $\tau_{ifc}$ is expected to be approximately independent of temperature.) $\kappa_c$ slowly increases with decreasing temperature. Since $c_{acoustic}$ is either saturated or decreasing with decreasing temperature and $v_c$ is expected to be only weakly temperature dependent, Eqtn. (6) implies that $\lambda$ also increases with decreasing temperature, as would be expected for phonon-phonon scattering. This is again consistent with $\lambda$ being larger than the interlayer spacing.

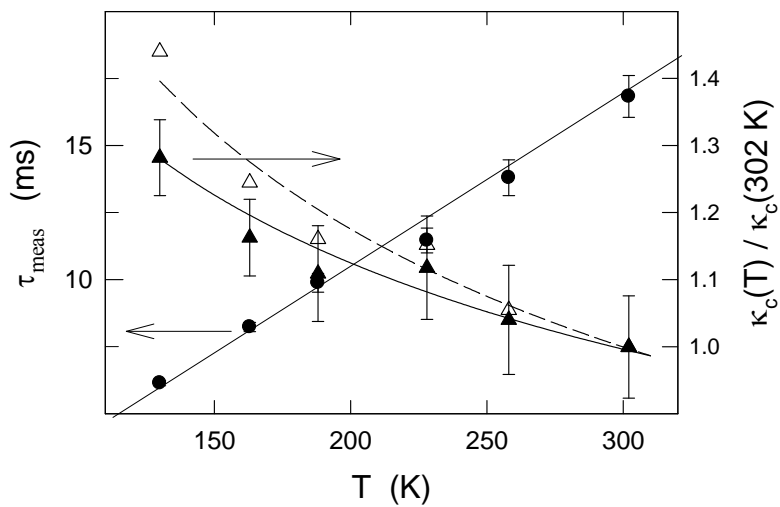

**Figure 4.** The temperature dependence of the measured thermal time constant and transverse thermal conductivity, normalized to its room temperature value, for rubrene, determined from $fV_\omega$ for crystal 2. The solid triangles represent the values of $\kappa_c$ assuming $\tau_2 = \tau_{meas}$, with error bars determined from the uncertainties in specific heat and $\tau_{meas}$. The open triangles are the values of $\kappa_c$ determined assuming a (temperature independent) interface time constant $\tau_{ifc} = 1$ ms. The curves are guides to the eye.

In summary, we have discussed the use of ac-calorimetry to measure the transverse thermal diffusivity of thin samples of poor thermal thermal conductors. We have used thin thermocouple wires to measure the temperature oscillations; a major constraint is in the thermal time constant of the glue used to attach the thermocouple to the sample. For rubrene, we have found that the interlayer thermal conductivity, $\kappa_c < 1$ mW/cm·K, is much smaller than the in-plane value, indicating poor thermal transport through phenyl side groups of the molecule. Nonetheless, the temperature dependence of $\kappa_c$ suggests that the interlayer phonon mean free path must still be at least a few lattice constants.

We thank Vitaly Podzorov and Yuanzhen Chen for providing rubrene samples, John Anthony for providing TIPS-pn samples, and Emily Bittle, Xiunu Lin, and Yulong Yao for technical assistance. This research was supported in part by the U.S. National Science Foundation through Grant Nos. DMR-0800367 and EPS-0814194 and the Center for Advanced Materials.